\documentclass{ragtime}
\usepackage{times}
\usepackage{color}  
\title[Constraints on two Kerr \textbf{BH} accretion tori]%
 {Constraints on two accretion disks\\ centered on the equatorial plane\\ of a~Kerr SMBH}

\author[D. Pugliese and 
 Z. Stuchl\'{\i}k]
 {Daniela Pugliese\at[]{1,a} 
 and Zden\v{e}k Stuchl\'{\i}k\at[]{1,b}\\
 \ins{1}Institute of Physics and Research Centre of Theoretical Physics and Astrophysics,\splitins[1]
 Faculty of Philosophy \& Science,
 Silesian University in Opava,\splitins[1]
 Bezru\v{c}ovo n\'{a}m\v{e}st\'{i} 13, CZ-74601 Opava, Czech Republic
 \\
 \ins{a}\Email{d.pugliese.physics@gmail.com}\\
 \ins{b}\Email{zdenek.stuchlik@physics.cz}}

\def\beq{\begin{equation}}
\def\eeq{\end{equation}}
\def\bea{\begin{eqnarray}}
\def\eea{\end{eqnarray}}

\newcommand{\il}{~}

\newcommand{\cc}{\mathrm{C}}

\graphicspath{{pug/}}

\begin{document}

\begin{abstract}
The possibility that two toroidal accretion configurations may be orbiting around a~super--massive Kerr black hole has been addressed. Such tori may be formed during different stages of the Kerr attractor accretion history.
We consider the relative rotation of the tori and the corotation or counterrotation of a~single torus with respect to the Kerr attractor. We give classification of the couples of accreting and non--accreting tori in dependence on the Kerr black hole dimensionless spin. We demonstrate that only in few cases a~double accretion tori system may be formed under specific conditions.
\end{abstract}


\section{Introduction}\label{intro}
Investigation of the issues concerning the attractor--accretion disks systems has led to question, if several toroidal orbiting structures may be formed around a~single central super--massive attractor. The debate on the possible evidence of their existence eventually ended in the conjecture that the occurrence of the unstable phases of these structures may be important in the high--energy phenomena demonstrated, for example, in active galactic nuclei ({AGNs}). This possibility has been addressed over the years with different methods and considering different astrophysical contexts. Possible observational evidences of these configurations were already discussed in \cite{2010ApJ...725.1507K,2011MNRAS.418..276S}.
Tori, strongly misaligned with respect to the central super--massive black hole (\textbf{BH}) spin, are considered in
\cite{2012ApJ...757L..24N,2013MNRAS.434.1946N,2014MNRAS.445.2285D},
where configurations around super--massive \textbf{BH} binaries are also addressed.
Toroidal fluid configurations, {tori},  might be formed as remnants of several accretion regimes occurred in various phases of the \textbf{BH} life {\citep{2013ApJ...771..119A,2007MNRAS.377L..25K,2006MNRAS.373L..90K,2002ApJ...573..754K}}. These sub--structures could be eventually reanimated in non isolated systems where the central attractor is interacting with the environment, or in some kinds of binary systems. Some additional matter could be supplied into the vicinity of the central black hole due to tidal distortion of a~star\citep{2015Natur.526..542M}, or if some cloud of interstellar matter is captured by the strong gravity. This issue has been faced in \cite{2015PhRvD..91h3011P,ringed} and then in \cite{dsystem} in the framework of the ``ringed accretion disk'' (\textbf{RAD}). It has been directly considered that in modelling the evolution of super--massive \textbf{BH} in {AGNs}, both corotating and counterrotating accretion stages are mixed during various accretion periods \cite{2003ApJ...582..559V,2015MNRAS.453.1608C,2015MNRAS.446..613D}, tidally destroyed stars corotating or counterrotating with respect to the attractor could leave some remnants in the form of toroidal structures which then can give rise to individual accretion tori. The tori can be accreting or non--accreting (i.e. isolated in equilibrium state)\footnote{Note that in Kerr--de~Sitter spacetimes torii can be also excreting\citep{2000A&A...363..425S,2005CQGra..22.3623S,2005MPLA...20..561S,2009CQGra..26u5013S}.}.
By following general arguments implied by the geometric properties of the Kerr spacetimes, in the present analysis we discuss the existence of couples of toroidal configurations, identifying the situations where a~doubled system of accretion tori may be formed in dependence on the \textbf{BH} spin. We consider couples of tori which are axis--symmetric and coplanar with the central Kerr \textbf{BH} as it is the simplest scenario adopted in the majority of the current analytical and numerical models of accretion configurations.
We proved that the existence of couples of accretion tori orbiting around a~central Kerr \textbf{BH} is strongly constrained and eventually only few coupled accretion tori systems are possible.
We also identify the Kerr attractors around which such tori may be formed.

Finally Sec.\il(\ref{Sec:app-graphs}) concerns with the construction and interpretation of special structures, graphs, representative of a~couple of accretion tori and their evolution, within the constraints they are subjected to. These structures were introduced in \cite{2016ApJS..223...27P} and detailed in~\cite{dsystem}.
Although the following analysis has been discussed independently from the graph formalism, graphs can be used also to quickly collect the different constraints on the tori existence and evolution. The use of these graphic schemes has been reveled to be crucial for the study and representation of the tori evolutionary paths.

\section{Kerr geometry as governor of doubled tori system}\label{Sec:Kerr-go}
The toroidal configurations are governed by the Kerr geometry through the radial profile of the specific angular momentum of the circular geodesic motion
\beq
\ell_{\pm}(r;a)\equiv-\frac{p_{\phi}}{{p_t}}=\frac{a^3M +aMr(3r-4M)\pm\sqrt{Mr^3 \left[a^2+(r-2M)r\right]^2}}{[Ma^2-(r-2M)^2r]M}
\eeq
defined as ratio of the two constants of test particle motion, $(p_{\phi}, p_t)$,
associated to the axial Killing field $\xi_{\phi}$, and the Killing field $\xi_t$ representing the stationarity of the Kerr geometry that is expressed in the Boyer--Lindquist coordinates $(\{t, r, \theta, \phi\})$, where $M$ denotes the black hole mass parameter, and $a=J/M$ is the specific angular momentum of the black hole having intrinsic angular momentum $J$--\citep{ringed}. The parameter $a/M=J/M^2\in[0,1]$ is the dimensionless spin of the black hole -- \citep{2013EL....10119001P,2013MNRAS.428..952P}.
The relative properties of the functions $\ell_{+}(r;a)$ for the counterrotating and $\ell_{-}(r;a)$ for the corotating motions govern the possible existence of the double toroidal configurations \citep{ringed}.

We consider two axially symmetric toroidal disks with symmetry plane coinciding with the equatorial plane of the central Kerr black hole.
As fast (slow) attractors we will mean black holes with high (low) spin--mass ratio with respect to some limits provided on $a/M$ by this analysis.
The accretion tori corotate or counterrotate with respect to the central Kerr \textbf{BH}, for $\ell_- a>0$ or $\ell_+ a<0$, if $\ell_{\pm}$ is for example the fluid constant specific angular momentum. In the following we drop the notation $\mp$ when it will be not necessary to specify the fluid rotation.
Considering the case of two orbiting tori, indicated by $(i)$ and $(o)$ respectively, we need to introduce the concept of \emph{$\ell$corotating} (\textbf{$\ell$c}) tori, defined by the condition $\ell_{i}\ell_{o}>0$, and \emph{$\ell$counterrotating} (\textbf{$\ell$r}) tori, defined by the relations $\ell_{i}\ell_{o}<0$. The two (\textbf{$\ell$c}) couples can be both corotating, $\ell a>0$, or counterrotating, $\ell a<0$, with respect to the central Kerr attractor.

\begin{figure}
	\includegraphics[width=.7\hsize,angle=90]{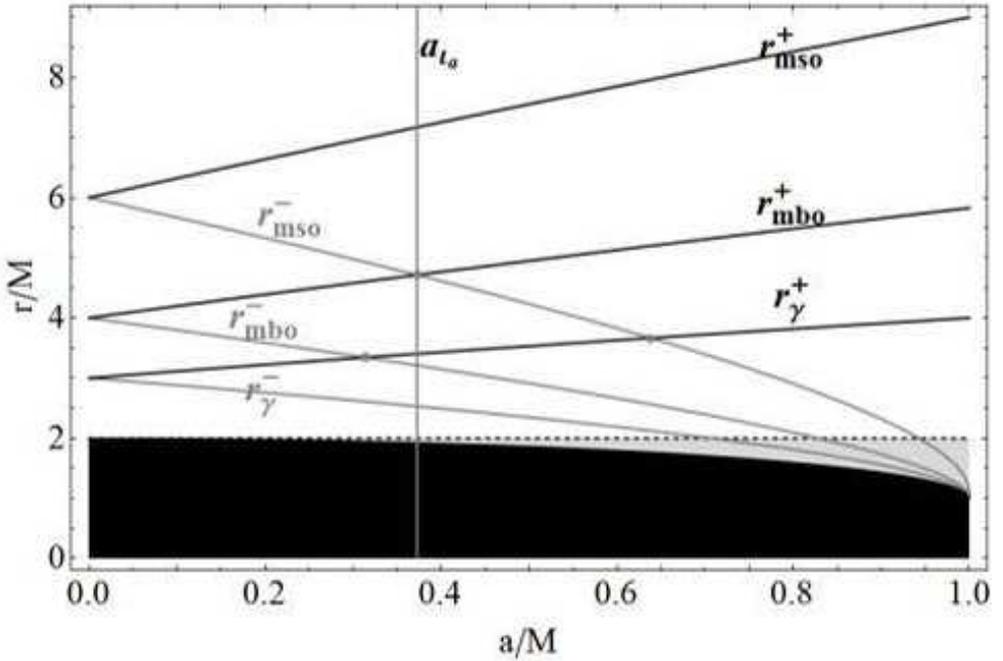}
	\caption{\textbf{Geodesic structure of the Kerr geometry on the equatorial plane $\theta=\pi/2$.} Radii $R_{\mathbf{N}}^{\pm}=\{r_{\gamma}^{\pm},r_{mbo}^{\pm},r_{mso}^{\pm}\}$ as function of the dimensionless spin $a/M\in[0,1]$ of the central Kerr attractor: the {marginally stable circular orbit} $r_{mso}^{\pm}$, the \emph{marginally bounded circular orbit} $r_{mbo}^{\pm}$ and finally the {marginal (photon) circular orbit} $r_{\gamma}^{\pm}$, for counterrotating (black curves) and corotating (gray curves) motion respectively. Black region is $r<r_+$, where $r_{+}$ is the outer Kerr horizon, gray region is the ergoregion or $r<r_{\epsilon}^+$, where $r_{\epsilon}^+=2M$ (dashed line) is the outer ergosurface on the equatorial plane of the Kerr geometry.	\label{PlotdisolMsMb}}
\end{figure}

In the study of the accretion disk dynamics in the Kerr geometry, and in particular here for the (\textbf{$\ell$r}) couples, it is important to consider the special radii, $R_{\mathbf{N}}^{\pm}$, determined by the geodesic characteristics of the Kerr spacetimes: the \emph{marginally stable circular orbit}, $r_{mso}^{\pm}$, the \emph{marginally bounded circular orbit}, $r_{mbo}^{\pm}$ and finally the \emph{marginal circular orbit} $r_{\gamma}^{\pm}$ (which is also a~photon circular orbit) \citep{2013PhRvD..88b4042P,2011PhRvD..84d4030P,2015EPJC...75..234P} -- Fig.\il\ref{PlotdisolMsMb}.
\begin{figure}
	\includegraphics[width=\hsize]{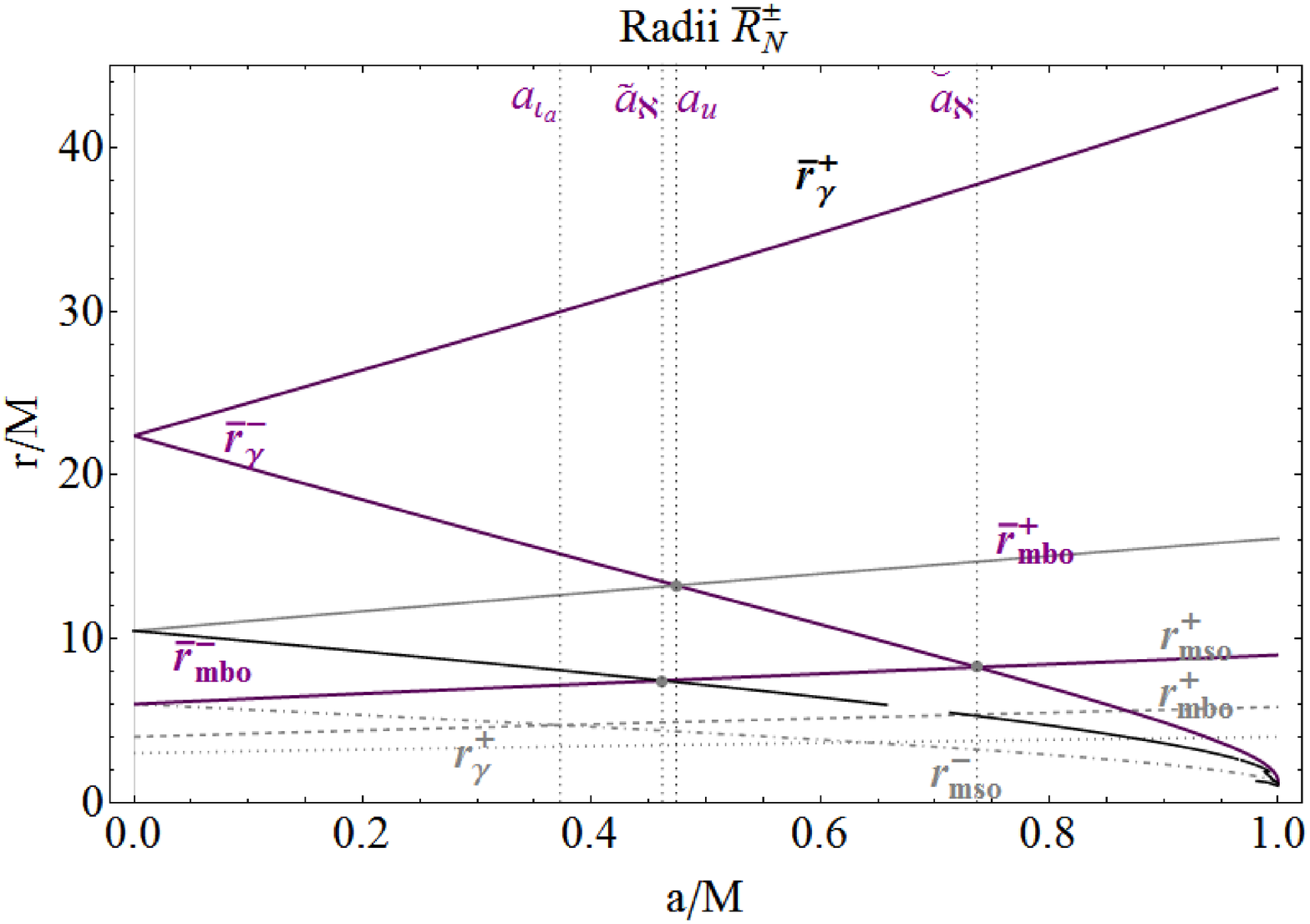}
	\caption{\textbf{Radii $\bar{R}_{\mathbf{N}}^{\pm}=\{\bar{r}_{\gamma}^{\pm},\bar{r}_{mbo}^{\pm},\bar{r}_{mso}^{\pm}\}$ } as function of the dimensionless spin $a/M\in[0,1]$ of the central Kerr attractor, where $r_{\epsilon}^+=2M$.	 \label{Fig:ConfCNN}}
\end{figure}

We assume \textbf{Condition 1}:

According to the symmetry conditions, the (stress) inner edge of the accretion torus is $r_{\times}\in]r_{mbo},r_{mso}]$, while the  torus is centered in $r_{\mathbf{\odot}}>r_{mso}$, according to the geodesic structure determined by the attractor and the torus corotation or counterrotation with respect to this.

\textbf{Condition 1} is consistent with most of the axially symmetric accretion torus models. Definition and constraints on the inner edge of the accretion torus and the  center may be found in \cite{2002ApJ...573..754K,1998Natur.391...54B,2010A&A...521A..15A}.
The double system of accretion tori is then strongly constrained by the geodesic structure of the Kerr spacetimes.
Solutions of the equations $r^{\pm}_{i}=r^{\mp}_j$ where $i\neq j$ and $r^{\pm}_{i}\in R_{\mathbf{N}}^{\pm}$ identify several notable spins $a_{\mathbf{N}}$ -- see Fig.\il\ref{PlotdisolMsMb} -- which distinguish several \textbf{BH} classes by some properties of the doubled torus system, as it will be discussed in \cite{multy}. Here we introduce the system of special radii $ \overline{R}_{\mathbf{N}}^{\pm}\equiv \{\overline{r}_{mbo}^{\pm},
\overline{r}_{\gamma}^{\pm}\}$
satisfying the equations $\ell_{mbo}^{\pm}\equiv\ell_{\pm}(r_{mbo}^{\pm})=\ell_{\pm}(\overline{r}_{mbo}^{\pm})$, in other words is the second solution of the equation $\ell(r; a)=\ell_{mbo}^{\pm}$, similarly  $\ell_{\gamma}^{\pm}\equiv \ell_{\pm}(r_{\gamma}^{\pm})=\ell_{\pm}(\overline{r}_{\gamma}^{\pm})$, which satisfies the relations $r_{\gamma}^{\pm}<r_{mbo}^{\pm}<r_{mso}^{\pm}<\overline{r}_{mbo}^{\pm}<\overline{r}_{\gamma}^{\pm}$--see Fig.\il(\ref{Fig:ConfCNN}).
Accordingly, we introduce the three characteristic values of the spin: \textbf{(1)}
$a_u\equiv0.4740M:$ $\overline{r}_{mbo}^+=\overline{r}_{\gamma}^-$, \textbf{(2)}
$ \tilde{a}_{\aleph}=0.461854M:$ $\overline{r}_{mbo}^-=r_{mso}^+$
and finally \textbf{(3)}
$ \breve{a}_{\aleph}\equiv 0.73688 M:$ $\overline{r}_{\gamma}^-=r_{mso}^+$, which govern the relations of $\overline{R}_{\mathbf{N}}^{\pm}$ and $R_{\mathbf{N}}^{\pm}$.

While, according to \textbf{Condition 1}, there is $r_{\times}\in]r_{mbo}^{\pm}, r_{mso}^{\pm}[$, then, as demonstrated in \cite{ringed,2016ApJS..223...27P} for an accreting torusthere is $r^{\pm}_{\odot}\in]r^{\pm}_{mso},\overline{r}^{\pm}_{mbo}[$. Whereas, {for any ``quiescent'' torus, i.e. non accreting torus, there is   $r_{\odot}>r_{mso}$, Figs\il\ref{Fig:teachplo},\ref{Fig:teachplocre}.}
More detailed constraints of the location of $r_{{\textbf{\mbox{$\mathbf{\times}$}}}}$ and $r_{\odot}$, particularly in the case of geometrically thick tori, are presented in \cite{ringed}.

The \textbf{Condition 1} and the relations among $\overline{R}_{\mathbf{N}}^{\pm}$ imply the following statements.

\section{Analysis of the double accretion disk system}\label{Sec:deal}
In the following, we present several statements that briefly represent results of our analysis of the doubled tori, while details on each of the presented cases will be provided elsewhere \citep{multy}.
We use the notation
$\cc^{\pm}$ for quiescent tori,
$\cc_{{\textbf{\mbox{$\mathbf{\times}$}}}}^{\pm}$ for accreting tori, then $\cc_{{\textbf{\mbox{$\mathbf{\times}$}}}}^+< \cc^-$ is the (\textbf{$\ell$r}) couple of an inner counterrotating accreting torus and an outer corotating quiescent torus represented in Figs\il\ref{Fig:teachplo}.
We refer in the following to the analysis of Sec.\il(\ref{Sec:Kerr-go}) and  particularly     the  \textbf{Condition I}.

Each torus is  modeled by a perfect fluid, geometrically thick  model, where the accretion phase (the start of accreting flow of matter towards the central \textbf{BH}) is driven by a the Paczynski-Wiita  (P-W) mechanism of destabilization of gravo-dynamical equilibrium. This is a mechanism of violation of mechanical equilibrium of the tori, i.e.  an
instability in the balance of the gravitational and inertial forces and the pressure gradients in the fluid \citep{abrafra}. The force balance in the tori is governed by the centrifugal, gravitational and pressure forces. A torus in accretion, $\cc_{\times}$, is therefore as in Figs\il(\ref{Fig:teachplo}): the  accretion onto the source occurs at the ``cusp''  $r_{\times}$ of the torus surface (the Boyer surface) where  the hydrostatic pressure is vanishing and the particles fall freely under the action of the gravitational field towards the attractor. The cusp $r_{\times}$   is therefore the inner margin of the accreting tori.
According to \textbf{Condition I} and   following discussion in Sec.\il(\ref{Sec:Kerr-go}), the position of the torus with respect to the central \textbf{BH}, is regulated by the range of variation of the fluid angular momentum, which evidently regulates the centrifugal component of the torus force balance.

 The larger is the fluid  specific momentum magnitude, the far away from the central \textbf{BH}  is the  torus (i.e., the torus center $r_{\odot}$).
Furthermore,  a larger  centrifugal  component in the force balance  acts in the sense to  prevent the accretion, i.e. accretion is  connected with a decrease in magnitude of  the angular momentum of the fluid. This holds also for a general rotational law where, for example, dissipative effects as viscosity, resistivity, and  the contribution of magnetic field are assumed. In the model adopted here for each torus, the centrifugal component  is balanced   with the gravitational and pressure components only.  As such, in order the accretion instability to occur,  the angular momentum magnitude has to be low ``enough'', namely $\mp\ell<\mp\ell_{mbo}^{\pm}$. We can also distinguish  the emergence of accretion (unstable point $r_{\times}$) from the emergence of  proto-jets, which are open cusped configurations not considered in this article--see \cite{open}.
 For the low fluids  angular momentum,  $\mp\ell^{\pm}\in]\mp\ell_{mbo}^{\pm},\mp\ell_{\gamma}^{\pm}[$, proto-jets appear as funnels of matter driven by the large centrifugal effects. Accretion occurs when $\ell\in[\ell_{mso},\ell_{mbo}]$, with proper values of $K$ parameter, which is related to the  effective potential function regulating the fluid equilibrium.
$K$ is constrained according to the  range of angular momentum. Because accretion occurs if  $K=K_{\max}<1$, where $K_{\max}$ corresponds to the maximum point  of the associated effective potential (explicit definition can be found for example in  \cite{abrafra,ringed}), and  the minimum point of the  hydrostatic pressure. The larger is $K$,   the bigger  is the torus (the  elongation in the equatorial plane).

 For   $r_{\odot}\gg r_{mso}$  there is $\ell>\ell_{\gamma}$ and the $K$ parameter, related to the fluid elongation is bounded in the range $K<1$, where $K=1$ is the asymptotic value  the effective potential. The gravo-hydrodynamic \textbf{P-W} instability cannot occur, and the Boyer surfaces are regular (absence of cusp)--see outer tori of Fig.\il(\ref{Fig:teachplocre}). This situation corresponds to the absence of a minimum pressure point defining the unstable point of the torus from the point of view of the force balance condition.
Concerning the location of the tori centers,  in the development of the \textbf{RAD} model in \cite{ringed}, radii  $\bar{R}_N $ have been introduced. These radii, defined  similarly to the set $R_N$, through the fluid angular momentum,   set the location of tori centers according to the ranges of specific angular momentum and reflect  the geodesic structure of the Kerr spacetime at greater distance.

Increasing the  magnitude of the fluid specific angular momentum,  the tori centers shift outwardly with respect to the black hole, the  $\bar{R}_{N}$ are the boundaries of the regions where the centers are located according to given $\ell$. Therefore, by considering $\bar{R}_{N}$, we can   more precisely restrict the position  of the torus center with respect to  the central attractor. These conditions reflect of the  geometrical structure of  the Kerr  spacetime and therefore they are describing  purely general  relativistic effect. Similar constraints are  considered also  in other models of accretion  tori  in presence of magnetic field or dissipative effects.

   The main idea of the analysis presented below consists in verifying that the condition on force balance  in the tori,  is fulfilled  for  a couple of tori to form and  for accretion from them to occur. Here we reduce the problem to the analysis of the set of parameters, $\ell$ and $K$,--details on this approach are thoroughly discussed in \cite{ringed,open,long,dsystem}.
   A great advantage of this approach is that this makes the torus analysis  simpler, providing an immediate way to fix constraints, and  eventually to provide  tori as initial configurations for more general GRMHD models.
   On the other side,  \emph{ranges} of variation for the specific  angular momentum of the fluids and the parameter $K$ are provided.  Finally, from the methodological point of view, it is clear that given the \textbf{Conditions I}, part of the analysis refers to the geodesic structure of Kerr's geometry, defined by the radii  $r^{\pm}_{i}\in R_{\mathbf{N}}^{\pm}$, limiting  the location of the critical points of the hydrostatic pressure in the torus (this holds also for other torus models where the edge of the accretion torus is  in the range  $[r_{mso},r_{mbo}]$). The angular momentum  fixes also the constraints on $K$ and therefore the outer  edge of the torus. As the possibility for accretion to occur is  regulated by the range of variation of fluid angular momentum, and the fluid angular momentum regulates  also the   location  with respect to the central \textbf{BH}, of the radii $R_{\mathbf{N}}^{\pm}$, then, we may set constrains for accretion  directly to the condition on   $R_{\mathbf{N}}^{\pm}$   rather then  to the angular momentum range. For an accurate description of how these procedures are applied  we refer to  \cite{dsystem}.  Further details, particularly on the role of the radii $R_{\mathbf{N}}^{\pm}$ could be found in  \cite{open},  while more discussion is in  \cite{long}.

\medskip

\begin{enumerate}
	\item\textbf{Quiescent tori } Two equilibrium (quiescent) tori, (\textbf{$\ell$c}) or (\textbf{$\ell$r}), can exist in the spacetime of any Kerr attractor, if their specific angular momenta are properly related \citep{multy}.
	\item
\textbf{The Schwarzschild attractor and the corotating accreting torus: constraints}
	If a~corotating toroidal disk is accreting onto the Kerr black hole, or if the attractor is \emph{static} ($a=0$), there \emph{cannot} exist any inner (corotating or counterrotating) torus between the outer accreting torus and the attractor.
	Then an outer, quiescent corotating torus can exist as is illustrated in Fig.\il\ref{Fig:teachplocre}--\emph{right} (\textbf{($\ell$c)}couples), or an outer counterrotating, quiescent or accreting disk can exist as illustrated in Fig.\il\ref{Fig:teachplo}--\emph{right}.
	\begin{figure}
		\includegraphics[width=\hsize]{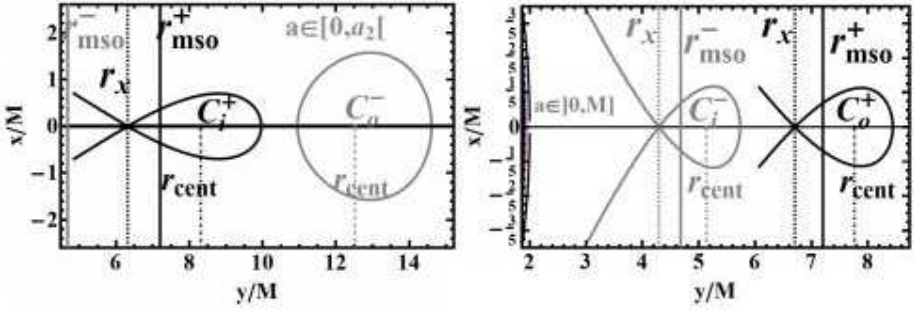}
	\caption{\textbf{$\ell$counterrotating couple} of accretion torus orbiting a~central Kerr black hole attractor with spin $a=0.382 M$. Cross sections on the equatorial plane of the  Roche lobes corresponding to  a~couple made by an inner counterrotating torus $\cc_{i}^+$ in accretion and an outer corotating torus $\cc_{o}^-$ (\emph{left}), and a~couple made by an inner corotating quiescent $\cc_{i}^-$ torus and an outer counterrotating torus $\cc_{o}^+$, being in accretion (\emph{right}). In couples $(\cc_i^+,\cc_o^-)$ with an inner torus in accretion, an outer torus may be formed around any Kerr attractor with $a\in[0,M]$ but only for $a\in[0,\tilde{a}_{\aleph}[$, the outer corotating torus of the couple may reach the condition for instability. Couples $(\cc_i^-,\cc_o^+)$ with one or both tori in accretion may exist around any Kerr attractor (excluding the Schwarzschild limit $a=0$ for a~static attractor). This illustrative case was obtained by integrating the equations for the Boyer surfaces in the general relativistic model of thick accretion tori in the Kerr geometry, with constant fluid specific angular momentum -- see for example \cite{ringed,2016ApJS..223...27P}: $r_{mso}^{\pm}$ are the marginally stable circular orbits for counterrotating and corotating matter respectively, $r_{{\textbf{\mbox{$\mathbf{\times}$}}}}\in]r_{mbo},r_{mso}[$ locates the inner edge of the accreting torus, $r_{mbo}$ is the marginally bounded orbit, $r_{\odot}$ is the center of the  correspondent Roche lobe (point of maximum hydrostatic pressure), and $(x, y)$ are Cartesian coordinates. For the case of a~couple ($\cc_{i}^+$, $\cc_{o}^-$) (\emph{left}) accretion phase is possible \emph{only} for the inner counterrotating torus (according to \textbf{Condition 1}). 	In the couple ($\cc_{i}^-$, $\cc_{o}^+$) (\emph{right}) the inner $\cc_{i}^-$ torus may accrete onto the attractor, or accretion may emerge from the outer counterrotating torus $\cc_{o}^+$, or both the tori can accrete, as shown in picture. Eventually in both cases, ($\cc_{i}^+$, $\cc_{o}^-$) and ($\cc_{i}^-$, $\cc_{o}^+$) collision between the  Roche lobes of the tori is in general possible.	 \label{Fig:teachplo}}
	\end{figure}
	\item
\textbf{Kerr attractor and  presence of a counterrotating accreting  torus.}
	If the accreting torus is \emph{counterrotating} with respect to the Kerr attractor, there is \emph{no} inner (the closest to the attractor) counterrotating torus between the attractor and the {outer} accreting torus, only outer quiescent corotating and counterrotating torus can exist as illustrated in Fig.\il\ref{Fig:teachplo}--\emph{left} and Fig.\il\ref{Fig:teachplocre}--\emph{left} respectively.
	\item
\textbf{ Location of the accreting  torus: the Schwarzschild case and the \textbf{ ({$\ell$c})} versus \textbf{ ({$\ell$r})}  cases.}
	For two \textbf{ ({$\ell$c})} tori, or if the attractor is static, two orbiting tori {cannot} \emph{both} be in accretion, neither the \emph{outer} torus of the couple may be in accretion, only the inner torus can be accreting as demonstrated in Fig.\il\ref{Fig:teachplocre}. In the \textbf{($\ell$r)} couple, an accreting corotating torus \emph{must be} the inner one of the couple. This implies that a~possible outer torus may be corotating (non accreting), or counterrotating (which may be in accretion under proper conditions) with respect the central black hole -- Fig.\il\ref{Fig:teachplo}.
	\item
\textbf{Focusing on the corotating torus of the couple.}
	A corotating torus can be the outer of a~couple where the inner accreting torus is counterrotating with respect to the Kerr black hole -- Fig.\il\ref{Fig:teachplo}--\emph{left}. In this case the outer corotating torus \emph{cannot be} accreting, for \emph{before} this occurs, the outer corotating torus starts to collide and eventually merges with the inner counterrotating torus.
	\begin{figure}
		\includegraphics[width=.9\hsize]{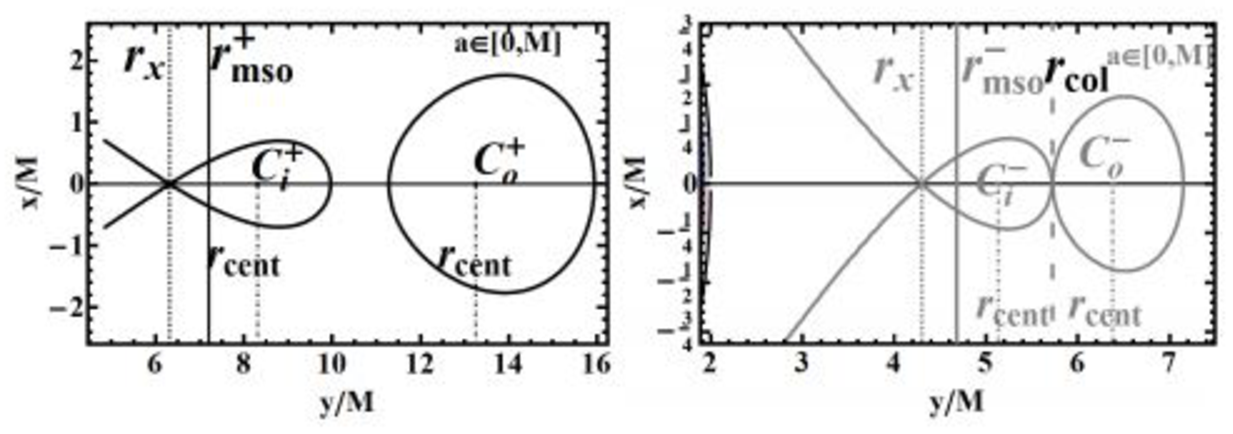}
		\caption{ \textbf{$\ell$corotating couple} of accretion torus orbiting a~central Kerr black hole attractor with spin $a=0.382 M$. Cross sections on the equatorial plane of the  Roche lobes correspondent to  a~couple of counterrotating tori $(\cc_{i}^+,\cc_{o}^+)$ (\emph{left}), and corotating tori $(\cc_{i}^-,\cc_{o}^-)$ (\emph{right}), $(x, y)$ are Cartesian coordinates. These couples with inner accreting tori may be formed around any Kerr black hole central attractor. Roche lobes are obtained as solutions for Boyer equipotential surfaces in the general relativistic model of thick accretion tori in the Kerr geometry, with constant fluid specific angular momentum -- see for example \cite{ringed,2016ApJS..223...27P}: $r_{mso}^{\pm}$ are the marginally stable circular orbits for counterrotating and corotating matter respectively, $r_{mbo}$ is the marginally bounded orbit, $r_{\odot}$ is the center of the  Roche lobe (point of maximum hydrostatic pressure), $r_{\textbf{\mbox{$\mathbf{\times}$}}}\in]r_{mbo},r_{mso}[$ locates the inner edge of the accreting torus. For an \textbf{($\ell$c)} couple accretion may emerge only from the inner torus. Collision between the  Roche lobes of the tori, here shown for the couple $(\cc_{i}^-,\cc_{o}^-)$ (\emph{right}), may be possible in any \textbf{($\ell$c)} couples.	\label{Fig:teachplocre}}
	\end{figure}
	\item \textbf{The couple made by an inner accreting counterrotating torus and an outer corotating torus.}
	The system consisting of an inner accreting counterrotating torus and an outer equilibrium corotating torus may be formed in any spacetime, but the {faster} is the attractor, the {farther away} should be the outer torus. This also implies the outer torus has large specific angular momentum \cite{ringed,dsystem,long,multy}.
	In this couple \emph{only} the inner counterrotating torus can accrete onto the attractor -- Fig.\il\ref{Fig:teachplo}--\emph{left}.
	We can particularize these arguments by saying that the corotating torus can be {``close''} to the phase of accretion (say here $r_{{\textbf{\mbox{$\mathbf{\times}$}}}}\gtrapprox r_{mso}^-$), before the merge may occur, only in the geometries of attractors with dimensionless spin $a\lesssim\tilde{a}_{\aleph}$. Collision between the tori may occur if $a\approx\tilde{a}_{\aleph}$ -- see \cite{dsystem}.
	\item \textbf{Double accretion I: couple $\cc_{\times}^-<\cc_{\times}^+$}
	If a~counterrotating torus
 is accreting onto the central black hole there could be an inner corotating torus, which may even accrete onto the spinning attractor -- see Fig.\il\ref{Fig:teachplo}-- \emph{right}. However, the lower is the Kerr black hole dimensionless spin, say $a\lessapprox a_u$,
	the lower must be the corotating torus specific angular momentum \citep{dsystem}.
	\item \textbf{Double accretion II}
	Remarkably, both the inner corotating torus of the couple and the outer counterrotating one of a~couple can accrete onto the attractor. In an (\textbf{$\ell$r}) couple, a~corotating torus can be accreting onto the attractor only if it is the \emph{inner} of the couple where the outer counterrotating torus can be in equilibrium or in accretion -- Fig.\il\ref{Fig:teachplo}--\emph{right}.
	A counterrotating torus can therefore reach the instability being the inner or the outer torus of an (\textbf{$\ell$r}) couple.
	\item \textbf{Summary: the situation for a torus in accretion}
	For a~torus in accretion the following two cases may occur: \textbf{ a)} for a~\emph{static} attractor only the \emph{inner} torus may accrete, while the outer of the couple has to be in equilibrium. A~torus collision occurs before the outer torus may accrete. This is the former point 2.
	\textbf{b)} if the torus is orbiting a~Kerr attractor then we need to distinguish the case of a~corotating torus from that of a~counterrotating one. \textbf{b-I} If the accreting torus is corotating with respect to the central Kerr black hole then the following two cases may occur: \textbf{b-I-i)} The accreting corotating torus is the inner torus of a~\textbf{($\ell$c)} couple, and the other corotating torus should be outer and in equilibrium -- Fig.\il\ref{Fig:teachplocre}--\emph{right}.
	\textbf{ b-I-ii)} The accreting corotating torus can be the \emph{inner} of an \textbf{($\ell$r)} couple where the outer torus is counterrotating and can be in equilibrium or in accretion -- Fig.\il\ref{Fig:teachplo}-\emph{right}.
	\textbf{b-II)} If, on the other hand, the accreting torus is counterrotating, three situations may arise. \textbf{ b-II-i)} The counterrotating accreting torus can be the inner of a~\textbf{($\ell$c)} couple where the outer one is in equilibrium, as follows from point 3 -- see Fig.\il\ref{Fig:teachplocre}--\emph{left}.
	\textbf{b-II-ii)} the counterrotating accreting torus may be the inner of an \textbf{($\ell$r)} couple where the outer one, corotating with respect to the black hole, has to be in equilibrium -- Fig.\il\ref{Fig:teachplo}--\emph{left}. The outer torus is close to the conditions for accretion only for slow attractors with
	$a<\tilde{a}_{\aleph}$.
	\textbf{b-II-iii)} The counterrotating accreting torus can be the \emph{outer} of an \textbf{($\ell$r)} couple where the inner corotating torus may be in equilibrium or in accretion -- Fig.\il\ref{Fig:teachplo}--\emph{right}.
	\item \textbf{Summary: the  \textbf{({$\ell$r})} couples}
	Focusing on the \textbf{({$\ell$r})} couples, suppose that a~\emph{counterrotating} torus will first be formed around a~Kerr attractor. This could be therefore the inner torus of a~possible couple. Then we know that \emph{only} the inner torus may accrete onto the attractor, no outer corotating or counterrotating torus can reach the instable phase in this double system.
	In the geometries of the fast Kerr attractors, with dimensionless spin $a\gtrapprox\breve{a}_{\aleph}$, the outer corotating torus of the \textbf{($\ell$r)} couple should be far enough from the attractor to avoid collision (for example for the perfect fluid thick torus model this would imply {$\ell>\ell_{\gamma}^-$ and $r_{\odot}>\overline{r}_{\gamma}^-$} -- \citep{2016ApJS..223...27P}). The slower is the attractor, the closest can be the outer corotating torus (at generally {lower specific angular momentum}).
	If, on the other hand, an {inner} \emph{corotating} torus will first be formed and it is accreting onto the central Kerr black hole,
	then an outer torus may be formed in any Kerr spacetime, it can be corotating or counterrotating. In this last case the outer torus can be also in accretion
	We have set out the range of location of the center $r_{\odot}$ through an assessment on the limits of variation of the specific angular momentum of the fluid, $\ell$, that has been especially studied for the geometrically thick torus model.
	This boosted precision in the limiting spins serves us only as an example of application of these results -- \citep{dsystem}.
	\item \textbf{Emergence of the instability in a couple}
	Considering the emergence of possible accretion phase, in all $a\in[0, M]$ geometries collision is possible under appropriate constraints, but for some cases collision is inevitable following the emergence of the unstable phases {of accretion} of one or both configurations of the couples.
	For $a\in[0, M]$ (\textbf{$\ell$c}) couples $\cc^{\pm}<\cc_{\times}^{\pm}$ cannot exist (merging with the outer equilibrium torus would precede the formation of this doubled system),
	while the (\textbf{$\ell$c}) couples, $\cc_{\times}^{\pm}<\cc^{\pm}$ and $\cc_{\times}^-<\cc^{\pm}$ may be formed.
	The (\textbf{$\ell$r}) couple $\cc_{\times}^{-}< \cc_{\times}^{+}$ is possible in $a\in]0, M]$:
	in the case of a~static attractor, described by the Schwarzschild geometry, there is $R_{\mathbf{N}}^+=R_{\mathbf{N}}^-$ and the \textbf{($\ell$r)} couples must fulfill the constraints of the (\textbf{$\ell$c}) couples.
	In the geometries $]\tilde{a}_{\aleph},M[$ only the
	(\textbf{$\ell$r}) couples where $\cc_{\times}^+ <\cc^-$ or $\cc^+ <\cc^-$, with $\cc^-$ close to accretion may exists .
	\item \textbf{Summary}
To summarize our analysis, we identify
	two classes of Kerr attractors, distinguished according to the features of the orbiting {\textbf{($\ell$r)}} couples: \textbf{a) } The fast attractors, with dimensionless spin $a\in]\tilde{a}_{\aleph},M]$ where there are no \textbf{($\ell$r)} couples, $\cc_{\times}^+ <\cc^-$ or $\cc^+ <\cc^-$, with an outer corotating torus close to the accretion phase -- point 6 and Fig.\il\ref{Fig:teachplo}--\emph{left}
	The {faster} is the Kerr attractor ($a\gtrapprox\breve{a}_{\aleph}$) the {farther away} should be the outer torus with large specific angular momentum to prevent collision -- point 9.
	The attractors with $a>a_u>\tilde{a_{\aleph}}$, where the outer counterrotating torus of the \textbf{($\ell$r)} couples $\mathrm{C}^- <\mathrm{C}_{\times}^+$ are possible in different conditions on the fluid specific angular momentum and the inner torus may be in accretion, as in Fig.\il\ref{Fig:teachplo}-\emph{right}, or quiescent -- see point 7.
	\textbf{b)}.
	The slow attractors, with dimensionless spin
	$a\in[0,\tilde{a}_{\aleph}[$: in these geometries an outer torus of a~$\cc_{\times}^+<\cc^-$ couple may be close to the accretion phase--Fig.\il\ref{Fig:teachplo}--\emph{left} point 6.
	In the field of attractors with $a<a_u $ the outer counterrotating torus may form a~$\cc^-< \cc_{\times}^+$ couple where the inner torus can be eventually in accretion -- see Fig.\il\ref{Fig:teachplo}--\emph{right}, but with sufficiently low specific angular momentum.
	Finally, Schwarzschild attractors ($a=0$) are characterized by \textbf{($\ell$r)} or \textbf{($\ell$c)} doubled systems where only the inner torus can be in accretion as illustrated in Figs\il\ref{Fig:teachplocre} or Figs\il\ref{Fig:teachplo}--\emph{left}. In fact, the geometric properties of the static spacetimes do not differentiate doubled systems according to the relative rotation of the two tori (point 2).
\end{enumerate}

\section{Conclusions}\label{conclus}

We proved that only specific couples of toroidal accretion tori may orbit around a~central Kerr black hole attractor: each torus is constrained by the presence of the second torus of the couple accordingly with the restriction provided by the Kerr background geometry.
Analogously, collisions between tori and then merging arise, especially in the emergency of the unstable phases of the single torus evolution, which leads ultimately to the accretion into the central Kerr black hole. As a~consequence of this the accretion of a~single torus would in fact be preceded by a~necessary merging with the torus companion. For all these reasons, the existence of the double torus systems appear to be in general strongly constrained and, ultimately, such a~configurations are possible only in few situations and under specific circumstances. Essential in this respect is the relative rotation of the tori in the couple, and the rotation with the respect to the central attractor.

The torus couples may be formed during different stages of the black hole life, interacting with the surrounding matter for tidal disruption of a~star or resulting from some gas clouds \citep{2012ApJ...757L..24N,2013MNRAS.434.1946N,2014MNRAS.445.2285D,2013ApJ...771..119A,2007MNRAS.377L..25K,2006MNRAS.373L..90K,2002ApJ...573..754K,2015Natur.526..542M}.

The setup provided in this letter provides the necessary conditions for the formation of these doubled configurations, and we expect they could be the starting point for further analysis in future investigations which may focus on their formation and especially the associated phenomenology.
The unstable phases of such configurations, we expect, may reveal of some significance for the high energy astrophysics related especially to accretion onto supermassive black holes, and the extremely energetic phenomena occurring in quasars and AGN that could be observable by the planned X--ray observatory ATHENA \footnote{http://the-athena-x-ray-observatory.eu/}.

\ack

D. P. acknowledges support from the Junior GACR grant of the Czech Science Foundation No:16-03564Y.

Z. S. acknowledges the Albert Einstein Centre for Gravitation and Astrophysics supported by grant No.
14-37086G.

\appendix

\section{Some notes on the graphs formalism}\label{Sec:app-graphs}
In this section we add some notes on the graph formalism introduced in \cite{2016ApJS..223...27P} and extensively used in \cite{dsystem,Letter,long,multy}.
The adoption of these schemes have proved to be an useful tool in the determination of the system states, as defined more precisely below, and especially in the investigation of tori evolutions as transition from different  states: from the initial equilibrium (non accretion) state to accretion or collision or, eventually, tori collision where with one or two tori in accretion.
Details concerning the graphs for a~tori couple of Figs\il\ref{Fig:teachplocre} and \ref{Fig:teachplo} have been extensively discussed in \cite{dsystem}, here we discuss some general aspects of the graph formalism clarifying some relevant notions.
As seen in Sec.\il(\ref{Sec:deal}), toroidal tori are related by boundary conditions dictated by the requirement of not penetration of matter (no tori overlapping) and by the geometric constraint for the equilibrium configurations determined by the geometric properties of the Kerr background.
These conditions however have to be relaxed to consider the case of tori collision in the macro--structure.
We distinguish four types of unstable couples of orbiting configurations (or unstable \emph{states} of the macro--configurations): \textbf{(I)} the
\emph{proto jet--proto jet} systems, corresponding to couples of open cusped surfaces, \textbf{(II)}
the \emph{proto jet--accretion} systems, where the proto--jet can follow or precede the accretion point, and finally the
\textbf{(III)} \emph{accretion--accretion} systems, where matter can accrete onto the attractor from several instability points.
However, not all these states can actually exist. We have proved that states formation and stability depend on the dimensionless spin of the attractor, the tori relative rotation respect to the central \textbf{BH}, and the relative rotation of the fluids in the tori (i.e. if they are $\ell$corotating or $\ell$counterrotating).

The five fundamental states, constituted by the $ \textbf{(I)} -- \textbf{(II)}$ and $ \textbf{(III)}$ couples constituted by all unstable configurations, and \textbf{(IV)} \emph{proto jet--non--accreting} configurations, \textbf{(V)} \emph{accreting --non--accreting} tori can be combined for \textbf{RADs} made up by more that two tori.
In this framework, the exploration of the internal dynamics of the macrostructure required the introduction of the notion of \emph{geometrical correlation} between two configurations of a~state, when the two surfaces may be in contact, in accordance with the constraints of the system \citep{2016ApJS..223...27P}, then feeding or collision phenomena happen, leading eventually to a~ transition of the couple state and, in the end, of the entire macro--configuration.
As a~consequence of this, we face the problem of the \emph{state evolution} i.e. the initial couple of configurations (starting state) evolves towards a~transition {from a stable solution to unstable one or viceversa}. This analysis has lead to conclusion for example that in some cases equilibrium configurations can only lead to proto--jet configurations and not to the accretion.

More specifically, two sub--configurations of a~\textbf{RAD} are said to be geometrically correlated if they may be in contact according to some constraints settled on their morphological or stability evolution.
A part of the \textbf{RAD} analysis is therefore dedicated to establish the possible geometric correlation of tori in the macrostructure giving a~number of features set in advance.
It is clear that a~geometrical correlation in a~ringed structure induces a~causal correlation in a~couple, when the morphology  of an element can be regarded as a~result of that correlation, in the instability such as collision.
On the other hand, causally correlated tori are tied by penetration of matter,
which implies necessarily a~geometric interaction between two sub--configurations. 

Graph schemes essentially serve to represent the evolution of a~couple, constituting the internal structure of a~\textbf{RAD}, thus below we introduce two important concepts regarding the accretion torus dynamics in a~state.

\textbf{(a)} the configuration sequentiality: the relative locations of the tori centers (the \emph{sequentiality} of the centers), \textbf{(b)} the relative location of the possible instability points (the \emph{sequentiality} of the maximum points of the effective potential).

\begin{figure}[h!]
		\includegraphics[width = 0.3\hsize]{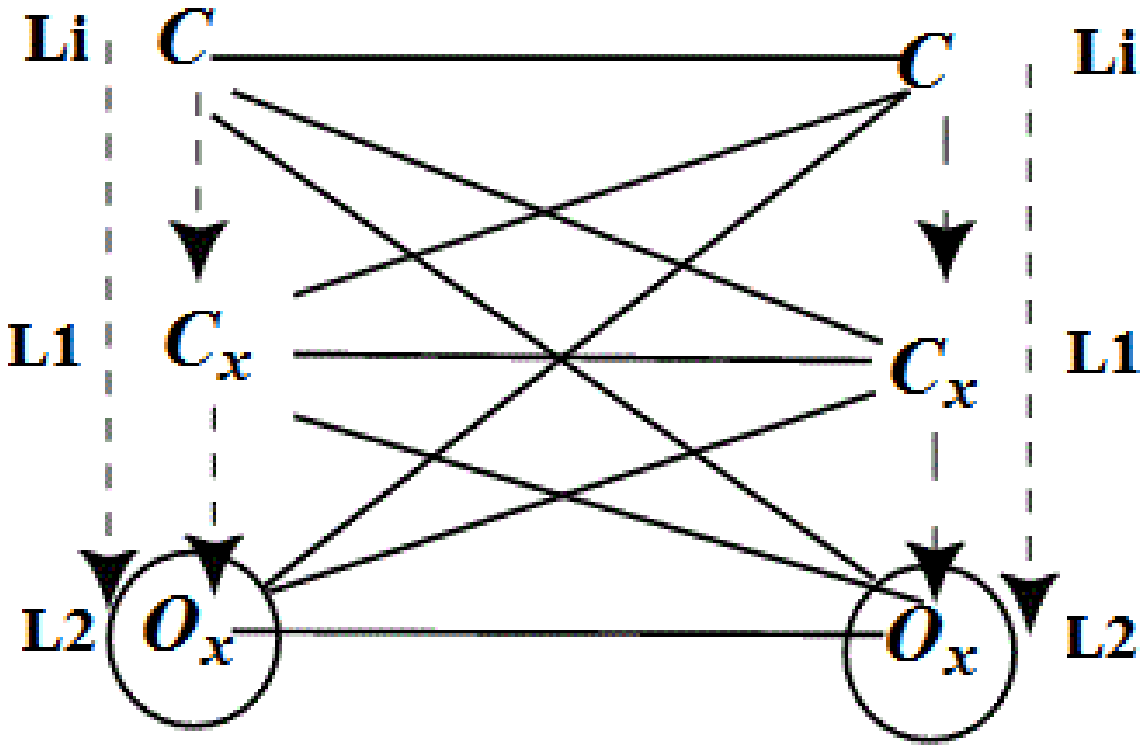}
		\includegraphics[width = 0.3\hsize]{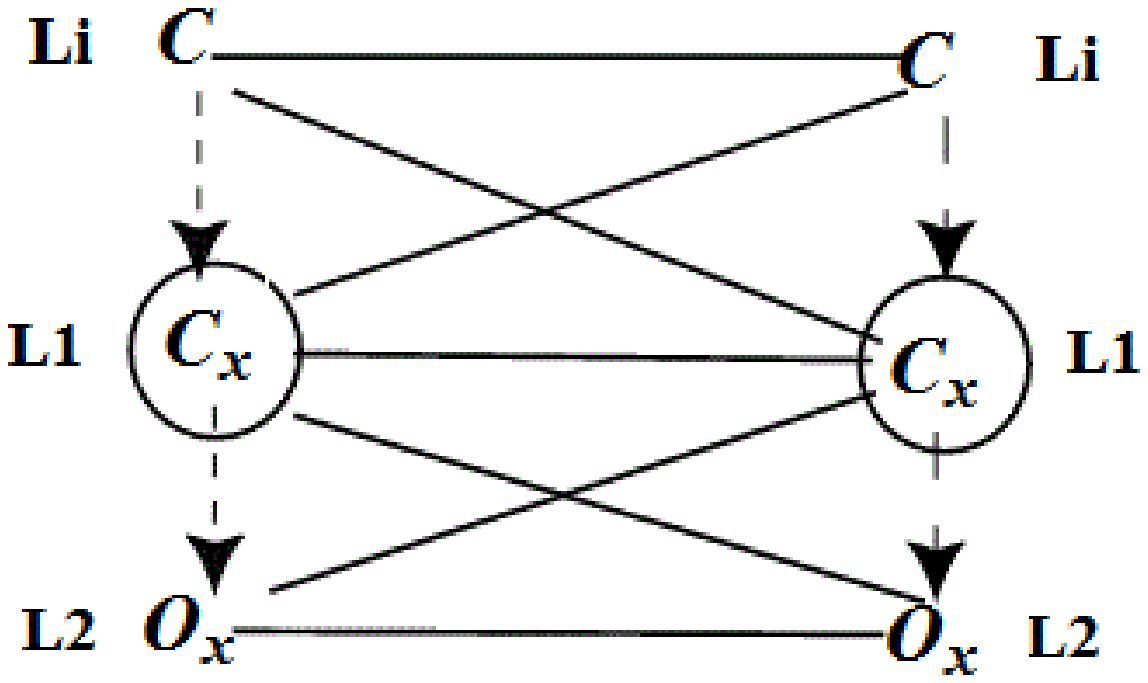}
		\includegraphics[width = 0.3\hsize]{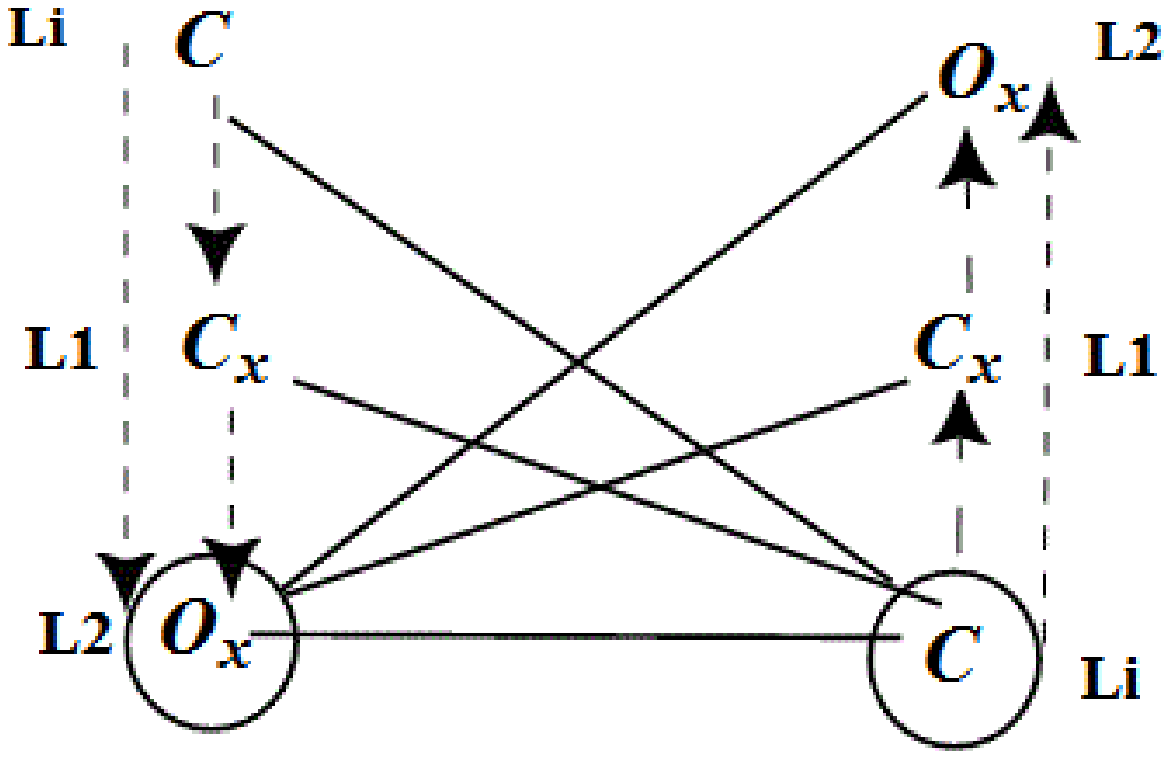}
		\includegraphics[width = 0.3\hsize]{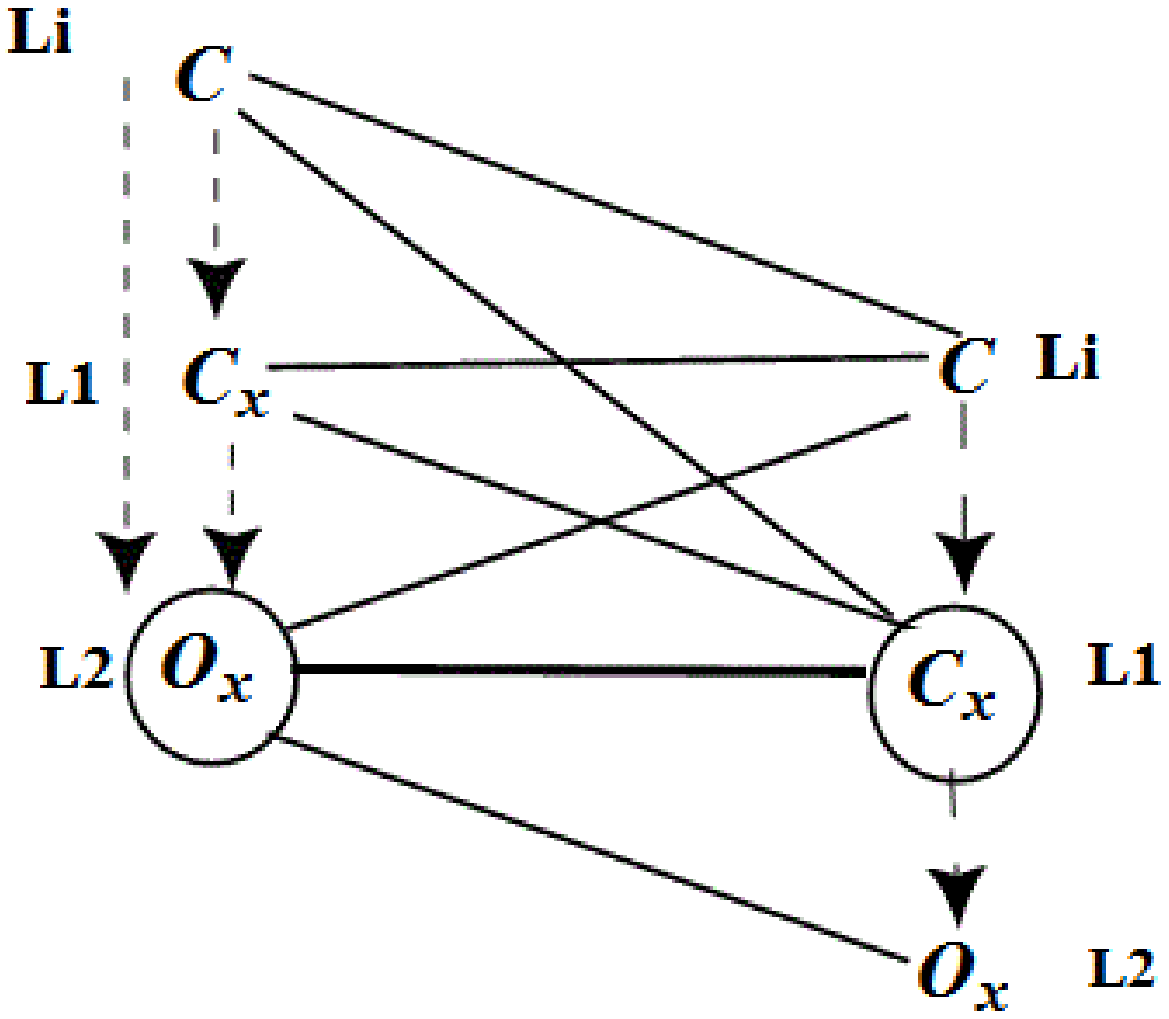}
		\includegraphics[width = 0.3\hsize]{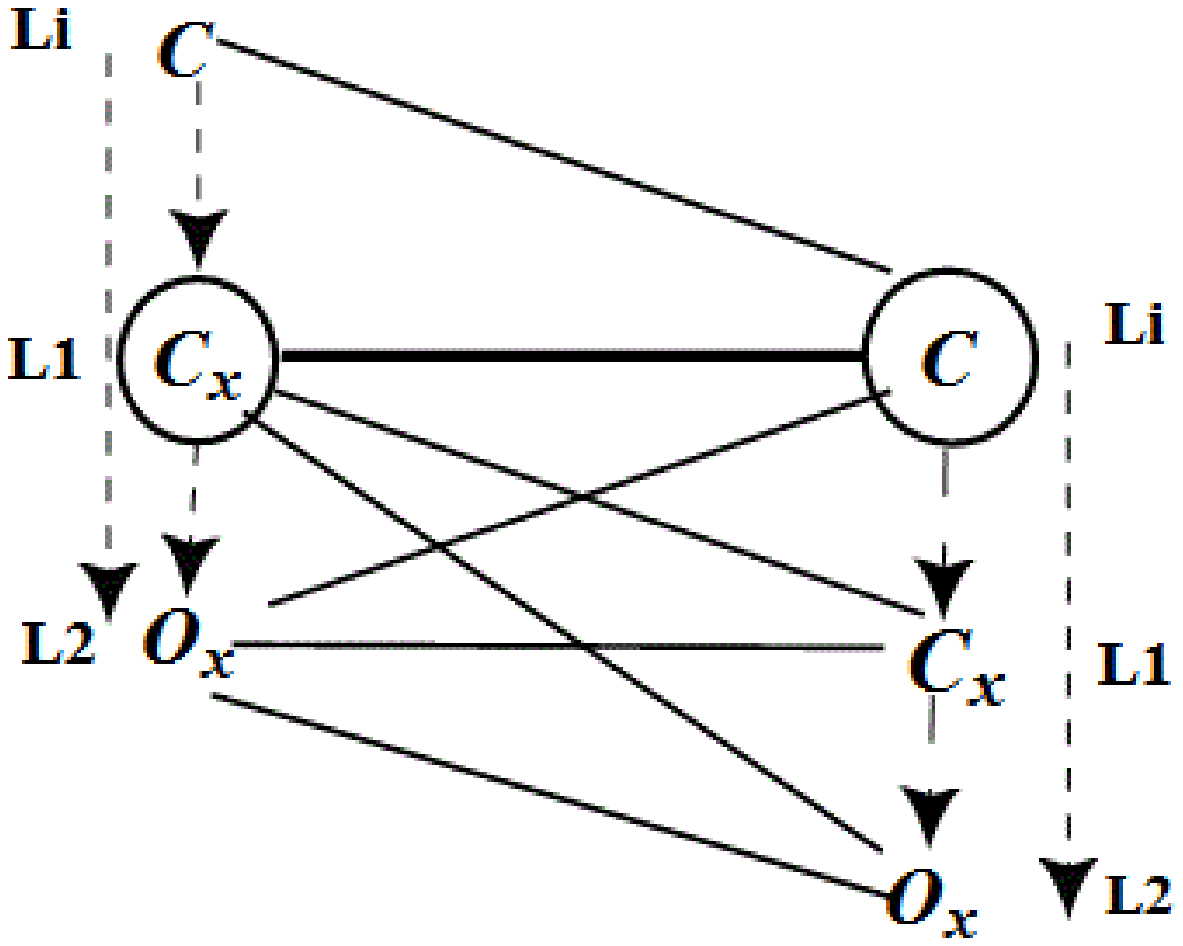}
	\caption{The five possible main states with at least one {unstable configuration}, for a~couple (circled) of the  classes $C$, $O$, $O_x$ and $C_x$. The stability is fixed
		but not the rotation with respect to the black hole, nor the sequentiality, according to the location of the maximum pressure points ($>$).
		Dashed arrow--lines (evolutive lines) show the evolution of the possible individual
		configuration, and start from the initial main state and point to the ending.
		Black continuum lines indicate the status of the couple
		in a~specific evolutionary phase of the couple. States with fixed sequentiality are represented by black lines with arrows, the direction of the arrow is in accordance to the sequentiality relation,	in the order of configuration (and criticality).
		Where uniquely fixed, the region of specific angular momentum is indicated close to each configuration. All the possible initial and final states of the evolutive lines to and from the circled couple	are shown.
		\label{Fig:COXOCX}}
\end{figure}

It is useful then to use the schemes in Fig.\il\ref{Fig:COXOCX}, where
five main states, with at least one {unstable (with the cusp)}, are sketched.
The different states for the
$\ell$corotating couples and the $\ell$counterrotating ones are specified in \cite{dsystem},
discussing the evolution of the possible individual
configurations, from the initial state to a~final one, following the \emph{evolutive lines} of Fig.\il\ref{Fig:COXOCX}, which actually represent transitions between (main) states of the couple, which are represented by the
\emph{state lines} of Fig.\il\ref{Fig:COXOCX}.
The analysis of \textbf{RAD} can be supported by
the construction of these diagrams and the determination of the constraints of the \emph{evolutive} and \emph{state lines}. Not all the initial and final states are possible and not all the evolutive lines are actually possible for different attractors.
Schemes of the main classes of Fig.\il\ref{Fig:COXOCX} can be seen as \emph{graphs} (see for example \cite{graphI}), 	each of the five main classes being therefore featured in the graphs.
	More precisely each scheme is a~vertex labeled graph that, we say, is \emph{centered} on a~main state (circle vertexes in Fig.\il\ref{Fig:COXOCX}) on which all the evolutive lines converge (point) or diverge. The evolutive lines are then oriented with the head and tail into two vertexes of the graph.
	There are two classes of lines: the evolutive lines which are oriented, and the state lines which are not oriented (in Fig.\il\ref{Fig:COXOCX}).
	Evolutive lines divide the graph into two parts centered around the center, namely in the \emph{antecedent section}, 	from which the heads of the arrows converging at the center start, and the \emph{subsequent section} which is the one onto which the evolutive lines 	starting from the center of the graph or crossing the center, with head in the antecedent section, converge, see Fig.\il\ref{Fig:COXOCX}.
	Two vertices can be connected by two evolutive lines pointing in the same direction (section), with different paths (however we assume throughout this work that the specific angular momentum and the $K$--parameter, whose variations determine the  transition {from a stable to an unstable solution}, do vary continuously).
	A graph may also have only one section, for example the first graph centered on the principal state of open configurations $ O_x-O_x$, where the only existing section is the antecedent one and then all the evolutive lines converge at the central state.
	A vertex can also belong to the states, and therefore vertexes, of two different sections. That is, it can be connected to vertexes in a~main state from which evolutive lines, having the opposite direction, start.
	However, an evolutionary line has only one possible orientation: an evolutionary loop that here means a~return to a~previous state for the same vertex is not considered {\citep{open,dsystem}}. As discussed in the text, this process may be possible, for example transition $C\dashrightarrow C_x\dashrightarrow C$. Two vertexes can be connected by two evolutive lines, say one is a~composition of (only) two evolutive lines connecting one vertex (the head) to a~third vertex and the second line connecting this vertex to the second one (the tail), which is a~composition of more lines representing different  steps. State lines are not oriented in the graphs (of the main states). A~state line has usually two possible ordinations, the sequentiality, of minimum or maximum respectively, then the lines are directed and undirected in this sense.
	We note that symmetry relations are not generally possible in the state lines, as in details discussed \cite{open,long}. For each center, the graph in Fig.\il\ref{Fig:COXOCX} is the one with the maximum number of vertices and lines.
	Then, we shall often eliminate some evolutive and the state lines of the graphs in Fig.\il\ref{Fig:COXOCX}, because they are not permitted for example in accretion tori around particular classes of attractors.
	The vertices are not completely labeled because there is missing the information on the specific angular momentum, in the case of equilibrium. We allow here to decorate a~graph, to specify the labels and other characteristics such as the line orientation.
	If all the possible information on the graph of a~main state are provided, for example when we have identified a~state element of the main state, the decorated graph is said saturated for complete decoration. {In some cases the state of a~system can remain indeterminate also after complete decoration -- Fig.\il\ref{Fig:COXOCX}}.
	All the vertices of the graphs are associated with lines. One can see the five graphs connected by state lines as a~\emph{macro--graph} connecting states. In fact, one can realize this just fixing up a~center of a~graph, and connect that to the other graphs looking for the same main state in the graphs centered in different states.
	All five graphs have degree or valency (i.e, number of vertices) six, vertex labels can be repeated for more vertex. The seed main states here have orders of composition $n=2$, no loops through state lines are possible. It is not excluded that such loops may be introduced when one use the five classes as a~seed of a~decomposition of higher order.
	A vertex can be associated to one and only one value of sequentiality according to the maximum points and the minimum points of the effective potential.
	Thus these graphs are simple, i.e., not oriented according to the sequentiality of the state lines, but they are not simple according to the evolutive lines.
	
	Eventually one can use the vertex colored graph notion combining two colors according to the signs of rotation relative to the attractor, a~different chromaticity associated with a~status line would correspond to a~$ \ell$counterrotating couple, an equal chromaticity to a~$ \ell$corotating couple.
	In this discussion, vertices connected or crossed by an evolutive line, pertain at an equal chromaticity. Thus if the center of the graph is monochrome then the entire graph is monochrome, and vice versa, if the center is dichromatic then the entire graph is dichromatic 	(intending the color of its vertices).
	Generally, we consider mixed graphs because we cannot always define the order to the state lines. On the other hand, since the different options of  evolution are marked by different evolutive lines, this uncertainty is not rendered in a~the mixed graph while remaining after the decoration. Instead, while not changing the stability during the evolution, one could think possible the evolutive line to change the sequentiality of the state line, but actually an evolutive line must represent transitions between monochromatic vertices ({configurations}), not changing, at fixed vertices, the sequentiality of the state line according to the minimum points of the effective potential, in fact this would imply an overlapping of matter and, similarly, in the evolution of a~monochromatic state the sequentiality according to the instability points is conserved , but it not necessarily conserved for a~dichromatic state as it is explained in \cite{ringed}. Two vertices are not always connected neither by an evolutive line or by a~state line, while the two sections of a~graph are always connected by evolutive lines crossing its center.

{To give an example  of the mechanism, we concentrate on Fig.\il(\ref{Fig:COXOCX})  picturing  a graph centered  on the couple   $\cc_{\times}^-<\cc{\times^+}$ featuring a double accretion. Here we specified the indication of the fluids  rotation and accordingly, the sequentiality, following the analysis in Sec.(\ref{Sec:deal}) which constraints a double accretion to  this specific couple only.
This graph has an antecedent   part
represented by the vertices of two quiescent tori. The initial state can be represented according to constraints on  fluids angular momentum and eventually the \textbf{BH} spin by any black line before the center,
 therefore the initial state may be formed by
two quiescent tori or one quiescent torus and an accreting  one of the center couple, different lines are for different sequentialities. Couple evolution  follows according to  the dashed lines, from one or two equilibrium tori to the state of two accreting tori. Eventually the center state may evolve towards the subsequent section of the graph, for a further transition. This  graph-case   may be found  as colored  graph in \cite{dsystem}.}


\def\prc{Phys. Rev. C }
\def\pre{Phys. Rev. E }
\def\prd{Phys. Rev. D }
\def\jcap{Journal of Cosmology and Astroparticle Physics }
\def\apss{Astrophysics and Space Science }
\def\mnras{Monthly Notices of the Royal Astronomical Society }
\def\apj{The Astrophysical Journal }
\def\aap{Astronomy and Astrophysics }
\def\actaa{Acta Astronomica }
\def\pasj{Publications of the Astronomical Society of Japan }
\def\apjl{Astrophysical Journal Letters }
\def\pasa{Publications Astronomical Society of Australia }
\def\nat{Nature }
\def\physrep{Physics Reports }
\def\araa{Annual Review of Astronomy and Astrophysics}
\def\apjs{The Astrophysical Journal Supplement}
\def\na{New Astronomy}

\def\mdash{---}

\bibliography{pug}

\end{document}